# Analyzing Unsolicited Internet Traffic: Measuring IoT Security Threats via Network Telescopes

Shereen Ismail*, Taelyn Dyer†, Raul Martinez†, Garrett Gastman†, Yozelyn Chavez†, and Asma Jodeiri Akbarfam†
* Merit Network, Inc., University of Michigan,
Ann Arbor, MI 48108, USA
† School of Engineering and Applied Sciences,
Washington State University Tri-Cities,
Richland, WA 99354, USA

***Abstract*—Network telescopes serve as a critical passive monitoring tool for capturing unsolicited Internet traffic, providing insights into global scanning and reconnaissance behavior. This study analyzes a 10-day dataset during January 2025 consisting of approximately 22 million packets collected by the ORION network telescope at Merit Network. By employing privacy-preserving metadata analysis and lightweight behavioral heuristics, we identify scanning and backscatter patterns without payload inspection. Our results reveal a highly structured and centralized ecosystem, where the top 1% of source IP addresses generate over 81% of total traffic. A significant finding is the dominance of Port 23 (Telnet) and Port 2323 (Telnet Alt), which highlights the persistent nature of IoT security threats and widespread attempts to exploit weak credentials in legacy IoT devices. Furthermore, synchronized surges in packet volume and Shannon entropy indicate coordinated, multi-vector reconnaissance campaigns. These findings offer a practical framework for identifying large-scale threat activity and support cybersecurity research and education.**

## I. Introduction

Unsolicited Internet traffic, often referred to as Internet Background Radiation (IBR), consists of packets sent to unused or unassigned IP addresses that did not initiate communication. This traffic is a critical signal for understanding global security trends, including botnet propagation and scanning campaigns. Network telescopes (aka darknets) capture this activity by monitoring unused but routable IP address space, providing a noise-free environment to study anomalous behaviors like service probing and backscatter [1], [2]. Despite their utility, operating these systems has become increasingly difficult; a recent survey identified only 28 distinct telescope initiatives, with just 18 believed to remain active [3], [4]. This decline is driven by the escalating value of IPv4 space, where a /19 block can be valued at nearly $395,673, leading major deployments like Merit's network telescope to reduce their address footprint by 60% since 2018 [3], [2].

This work has been submitted to the IEEE 7th World AI IoT Congress (AIIoT 2026).

The modern threat landscape is heavily shaped by the vulnerability of the IoT. Internet scanning has increased 30-fold over the last decade, growing from 11 million daily packets in 2015 to 345 million in 2024 [5]. Large-scale scanning often surges following major vulnerability disclosures, with attackers completing full IPv4 scans within 24–48 hours to recruit insecure IoT devices into botnets like Mirai [5]. Statistics from recent monitoring show that Port 23 (Telnet) consistently overwhelms other services, receiving over 20 million packets per month, a factor of four higher than other common ports [2]. This activity is fueled by automated tools and collaborative sharding that allow attackers to blanket the entire port range (all 65,536 ports) to discover alternate services on ports like 5555 (ADB) and 8080 (HTTP Alt) [5], [2].

This paper presents an empirical analysis of 1,000,000 connection records from the ORION network telescope at Merit Network, which represents one of the largest and most persistent darknet initiatives in the US, with a longitudinal history spanning nearly 20 years [6], [2]. We characterize traffic patterns across geographic regions, Autonomous Systems (ASNs), and destination services using lightweight heuristics. Our analysis demonstrates that unsolicited traffic is a highly structured phenomenon rather than random noise. We observe extreme traffic inequality, where the top 1% of source IP addresses generate over 81% of total activity. By correlating traffic surges with synchronized shifts in Shannon entropy, we provide a reproducible framework for identifying coordinated, multi-vector reconnaissance campaigns targeting the global IoT ecosystem. The main contributions of this paper are summarized as follows:

- We conduct an empirical analysis of unsolicited Internet traffic captured by the ORION network telescope, analyzing one million connection records collected over a ten-day observation period.
- We characterize large-scale scanning behavior across geographic regions, autonomous systems (ASNs), and destination ports using privacy-preserving metadata analysis.
- We demonstrate that unsolicited Internet traffic ex-

hibits extreme concentration, where a small fraction of source IP addresses generate the majority of observed traffic.
- We introduce an entropy-based temporal analysis that reveals synchronized increases in traffic volume and diversity, suggesting coordinated multi-vector reconnaissance campaigns targeting IoT-related services.
- We provide insights into high-risk service targeting patterns, particularly the persistent dominance of Telnet and web service ports, highlighting ongoing security risks in the global IoT ecosystem.

The remainder of this paper is organized as follows. Section II presents related work. Section III describes the dataset and methodology. Section IV presents the results and analysis. Section IV-G presents the discussion and future work. Section V concludes the paper and outlines directions for future work.

## II. Related Work

Network telescope research has evolved from simple traffic volume characterization to complex behavioral modeling and threat attribution. This section categorizes prior research according to the taxonomy of darknet research summarized in Figure 1, which distinguishes between data collection methods, analytical paradigms, and observed adversarial behaviors. A thematic comparison of significant literature mapping to these categories, including foundational trends, operational integrity, and behavioral analysis, is provided in Table I.

### A. Foundational Measurement and Longitudinal Trends

Longitudinal studies provide critical baselines for understanding IBR. Research from a Spain-based telescope characterized a year of traffic, identifying the dominance of TCP and DoS-related reflection patterns [7]. Long-term analysis of the Merit Network telescope spanning 20 years highlights evolving scanning trends and the impact of address space transitions [2]. Similarly, a decade-long study observed that modern scanning has become increasingly automated, targeting a concentrated set of common services [5].

### B. Operational Challenges and Infrastructure Scaling

Maintaining large-scale sensors faces mounting difficulties due to IPv4 address scarcity and escalating costs [3], [4]. While major deployments like the UCSD telescope have faced routing and packet loss distortions [4], research demonstrates that scaling down a telescope by half can still capture 80% of unique attack sources [3]. Reliability at the point of capture is also vital; studies comparing tools like DPDK and XDP show that specialized frameworks are necessary to handle high-volume surges without data loss [8].

### C. Behavioral Analysis and Machine Learning

To process massive datasets efficiently, researchers have moved toward metadata-centric analysis. The i-DarkVec framework uses representation learning to model interactions between sources and services as sequences, enabling the discovery of coordinated scanning campaigns [9]. Lightweight machine learning approaches utilize Layer 3 and 4 header features to classify scanning and backscatter events in near real-time without the overhead of payload inspection [1].

### D. Contextual Intelligence and IoT Security

Network telescopes are increasingly complemented by active sensors and external intelligence. Distinguishing between acknowledged research scanners and exploit-driven actors is essential for accurate threat assessment [10]. While telescopes provide a macro view of global IoT-centric scanning bursts, distributed honeypots offer micro-level insights by engaging with bots like Mirai to capture session semantics and credentials [11]. Active measurement systems further contextualize these observations by mapping the global landscape of exposed services [12].

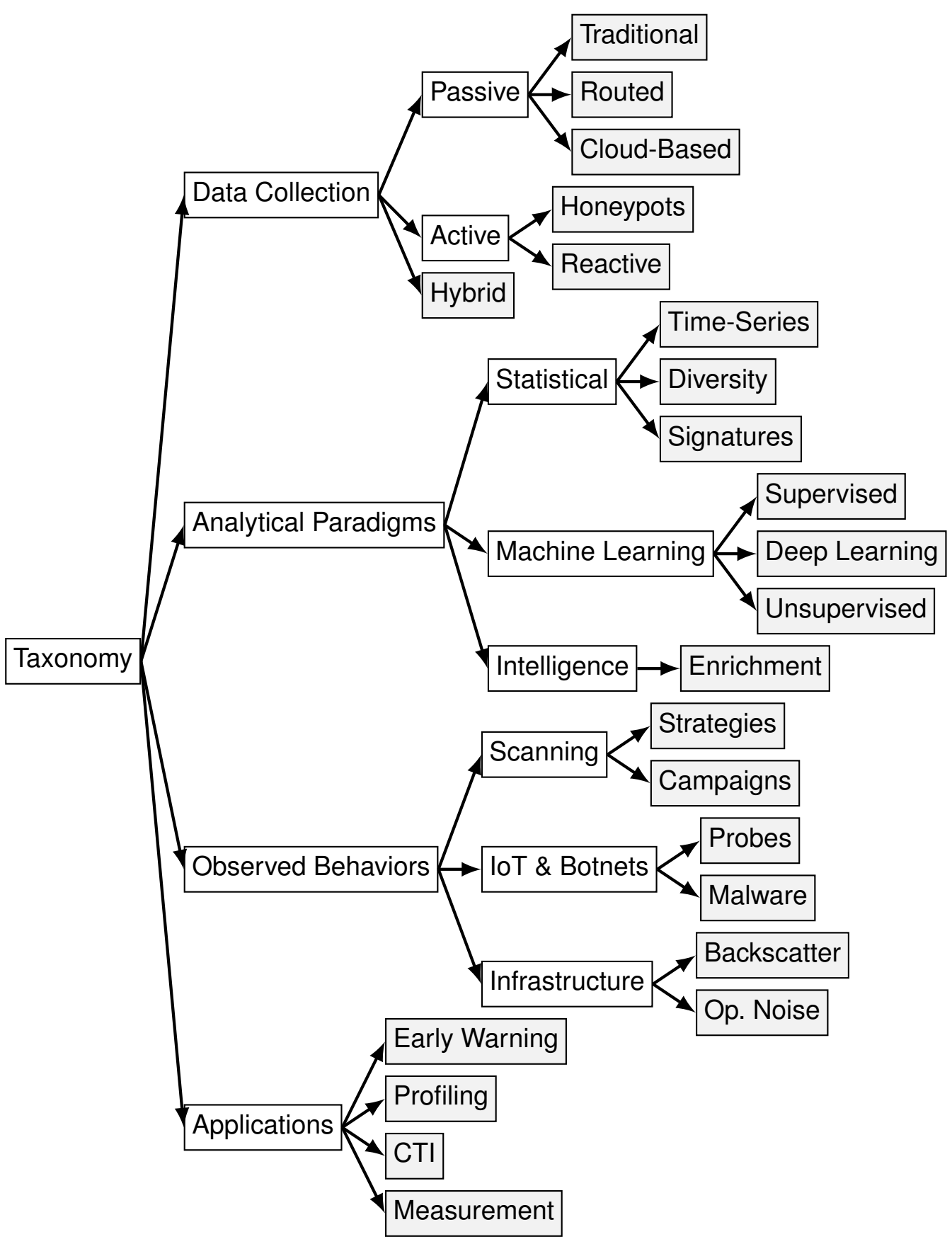

Fig. 1: Taxonomy of darknet research, categorized by collection, analysis, behavior, and applications

## III. Methodology

### A. Dataset Collection

We analyze unsolicited connection records collected by the ORION network telescope operated by Merit Network.

TABLE I: Thematic Comparison of Recent Network Telescope and IBR Research

| Study | Year | Category | Scale | Key Contribution |
|---|---|---|---|---|
| *Foundational & Longitudinal* | | | | |
| Griffioen et al. [5] | 2024 | Foundational | ∼/16 | Documents 10 years of scanning growth and extreme target volatility. |
| Ismail et al. [2] | 2025 | Foundational | /8, /13 | 20-year longitudinal analysis of Merit darknet and its /13 transition. |
| García-Peñas et al. [7] | 2025 | Foundational | /16 | Characterizes annual IBR trends and the persistent dominance of TCP scanning. |
| Yadav et al. [13]* | 2025 | Foundational | Survey | Unified taxonomy of scanning detection and adversary attribution. |
| *Operational & Scaling* | | | | |
| Männel et al. [4] | 2025 | Operational | /8, /9, /10 | Lessons from UCSD darknet on routing distortions and packet loss impacts. |
| Camargo et al. [3] | 2024 | Operational | /19 | Fidelity analysis proving halving sensor size retains 80% unique sources. |
| Ismail et al. [8] | 2025 | Operational | /13 | Benchmarks collection tool stability (XDP/DPDK) for high-volume IBR. |
| *Behavioral & ML Analysis* | | | | |
| Kallitsis et al. [14] | 2022 | Behavioral | /10 | Change-point detection using autoencoders and Wasserstein distance. |
| Gioacchini et al. [9] | 2023 | Behavioral | /24 | i-DarkVec incremental embeddings for clustering and characterization. |
| Ismail et al. [1] | 2025 | Behavioral | /13 | Lightweight ML for real-time scanning and backscatter classification. |
| Obaidat et al. [13] | 2025 | Behavioral | Survey | Review of IDPS strategies and advancements in adaptive supervised learning. |
| *Contextual & Intelligence* | | | | |
| Cabana et al. [15] | 2021 | Contextual | /13 | Generates ICS threat intelligence using payload clustering and entropy analysis. |
| Collins et al. [10] | 2023 | Contextual | /24 | Distinguishes acknowledged scanners from exploit-driven traffic. |
| Durumeric et al. [12] | 2015 | Contextual | IPv4 | Censys search engine for active service exposure and context analysis. |

*Refers to the survey portion of the bibliography provided in the attached sources.

The dataset consists of *connection-level summaries* rather than full packet captures, enabling analysis of scanning, misconfigurations, DoS patterns, and reconnaissance behaviors using metadata features instead of payload inspection. Each record includes fields such as `SourceIP`, `Port`, `Packets`, `Bytes`, and various behavioral indicators like `Zmap`, `Masscan`, and `Mirai`. An overview of the dataset source, collection site, and data representation is summarized in Table II.

### B. Preprocessing, Ethics, and Analysis Pipeline

Our preprocessing and analysis pipeline is designed to remain privacy-preserving while supporting security inference from Internet Background Radiation (IBR). Because the dataset provides aggregated connection records, no packet payloads are inspected. Analysis relies exclusively on metadata features such as ports, traffic volume, TCP indicators, and contextual attributes like country or ASN. The pipeline follows a structured multi-stage process (Figure 2):

- **Data Loading and Initial Parsing:** Raw CSV files were parsed using `pandas`. Column types were validated and cast. Records with NULL values in critical fields (`SourceIP`, `Port`, `Packets`) were excluded to ensure data integrity.
- **Data Cleaning and Filtering:** Records with zero packets or invalid ports were removed.
- **Feature Enrichment:** Geographic and network attributes were enriched using GeoLite2/geoip2, mapping source IPs to countries and ASNs. This enables both spatial analysis and ASN-level aggregation for behavioral insights.
- **Behavioral Heuristics:**
  1) **Scanning Detection:** Source IPs contacting five or more destination ports over the observed timeframe were flagged as potential scanners. This threshold provides a lightweight heuristic for distinguishing multi-port probing behavior from isolated connection attempts and helps reduce false positives caused by occasional or misconfigured connection attempts.
  2) **Backscatter Detection:** Records exhibiting TCP flag combinations consistent with victim responses to spoofed DoS attacks were identified.
  3) **Traffic Concentration Analysis:** Packet sums per ASN were computed to distinguish persistent versus bursty behavior.
- **Statistical Aggregation:** Data was aggregated by port, ASN, and geography to compute descriptive and inferential metrics. Traffic concentration was quantified using the Lorenz curve and Gini coefficient; entropy was computed hourly for port and ASN distributions to assess randomness and variability.

## IV. Results and Analysis

This section presents a multi-dimensional characterization of unsolicited traffic observed in the collected dataset. We analyze geographic concentration, ASN-level temporal dynamics, service-targeting behavior, traffic inequality, and entropy-based diversity. Together, these analyses reveal structural regularities consistent with automated scanning infrastructure, opportunistic reconnaissance, and coordinated large-scale probing activity.

### A. Geographic Distribution of Darknet Traffic

Figure 3 illustrates the geographic distribution of source countries generating unsolicited traffic. The United States contributes a substantially larger share than any other country, followed by a precipitous decline across remaining contributors such as Great Britain, China, Germany, and Japan. Other notable contributors include Malaysia, Singapore, Taiwan, Denmark, and Portugal.

TABLE II: Dataset Description and Detailed Attributes

| Attribute | Description |
|---|---|
| **Data Source** | ORION Network Telescope |
| **Collection Site** | Merit Network, Inc. |
| **Collection Period** | Jan 9, 2025 – Jan 19, 2025 (10 days) |
| **Data Format** | Connection-level records; pre-aggregated; CSV-style |
| **Data Volume (File Size)** | 334.1 MB (1,433,057,230 bytes) |
| **Number of Records** | 1,000,000 |
| **Total Number of Packets** | 22,026,154 |
| **Unique Source IPs** | 65,529 |
| **Unique Dest Ports** | 39,174 |
| **Unique ASNs** | 3,312 |
| **Temporal Resolution** | Hourly aggregation (for time-series plots) |
| **Data Fields (per Record)** | SourceIP, Port, Traffic, First, Last, Packets, Bytes, UniqueDests, UniqueDest24s, Lat, Long, Country, City, ASN, Org, Prefix, RDNS, Zmap, Masscan, Mirai, Samples, TCP, ICMP |
| **Data Parsing Method** | Records loaded via Python's `pandas`; validation of column types; removal of NULLs and invalid ports [cite: 10, 11, 12, 13] |

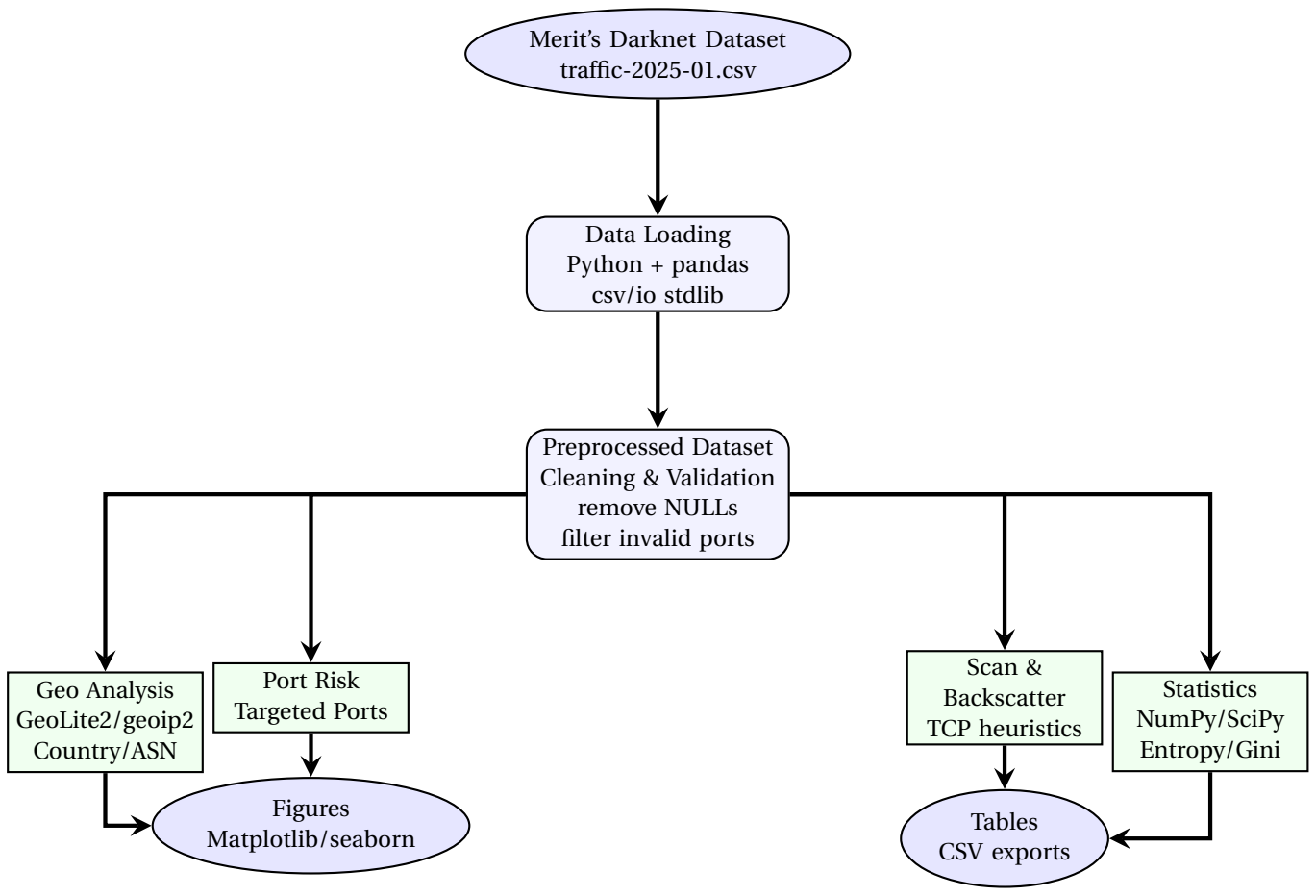

Fig. 2: Analysis workflow for the darknet dataset including preprocessing, enrichment, behavioral detection, and statistical analysis. Tools used at each stage are included within the workflow boxes.

This concentration likely reflects the centralization of global IT infrastructure, specifically the high density of cloud service providers (CSPs) and data centers in these regions. It is critical to distinguish between attribution and origin; geographic dominance indicates where scanning infrastructure, whether legitimate (e.g., Censys, Shadowserver) or malicious (e.g., botnet C2 nodes), is hosted, rather than the physical location of the threat actors. The sharp rank-order decay suggests that unsolicited activity is not a uniform global phenomenon but is rooted in specific, well-connected network ecosystems.

### *B. Temporal Traffic Patterns by ASN*

The temporal dynamics of the ORION dataset, illustrated in Figure 4, reveal a landscape defined by both steady-state reconnaissance and aggressive, short-lived surges. Analysis of the top five ASNs identifies three distinct behavioral signatures: **Persistent Scanning:** Skynet Network Ltd (AS214295) serves as the most consistent contributor, maintaining a steady baseline of activity between $10^4$ and $10^5$ packets from January 17 through the end of the window. Such a profile is characteristic of dedicated, automated scanning infrastructure performing continuous Internet-wide reconnaissance. **Discrete Spikes:** Conectec Net Ltda (AS269538) generated a massive, isolated spike exceeding $5 \times 10^6$ packets on January 16. Similarly, Mevspace sp. z o.o. (AS201814) contributed in disjointed intervals, appearing prominently during the final surge on January 19. These patterns suggest episodic activity, such as short-lived vulnerability research or transient botnet coordination. **Coordinated Terminal Surge:** A pronounced escalation in activity occurs late on January 19, where the total traffic (indicated by the dashed line) surges toward $10^7$ packets. This event is driven by the simultaneous convergence of activity from Google Cloud Platform (AS396982), Microsoft Corporation (AS8075), and the sustained output of Skynet Network Ltd. The synchronized nature of this final spike indicates a distributed probing event where multiple distinct network origins, ranging from major cloud providers to specialized hosting services, targeted the telescope simultaneously. This transition from sporadic background noise to a multi-source coordinated surge underscores the highly structured nature of modern, large-scale reconnaissance ecosystems.

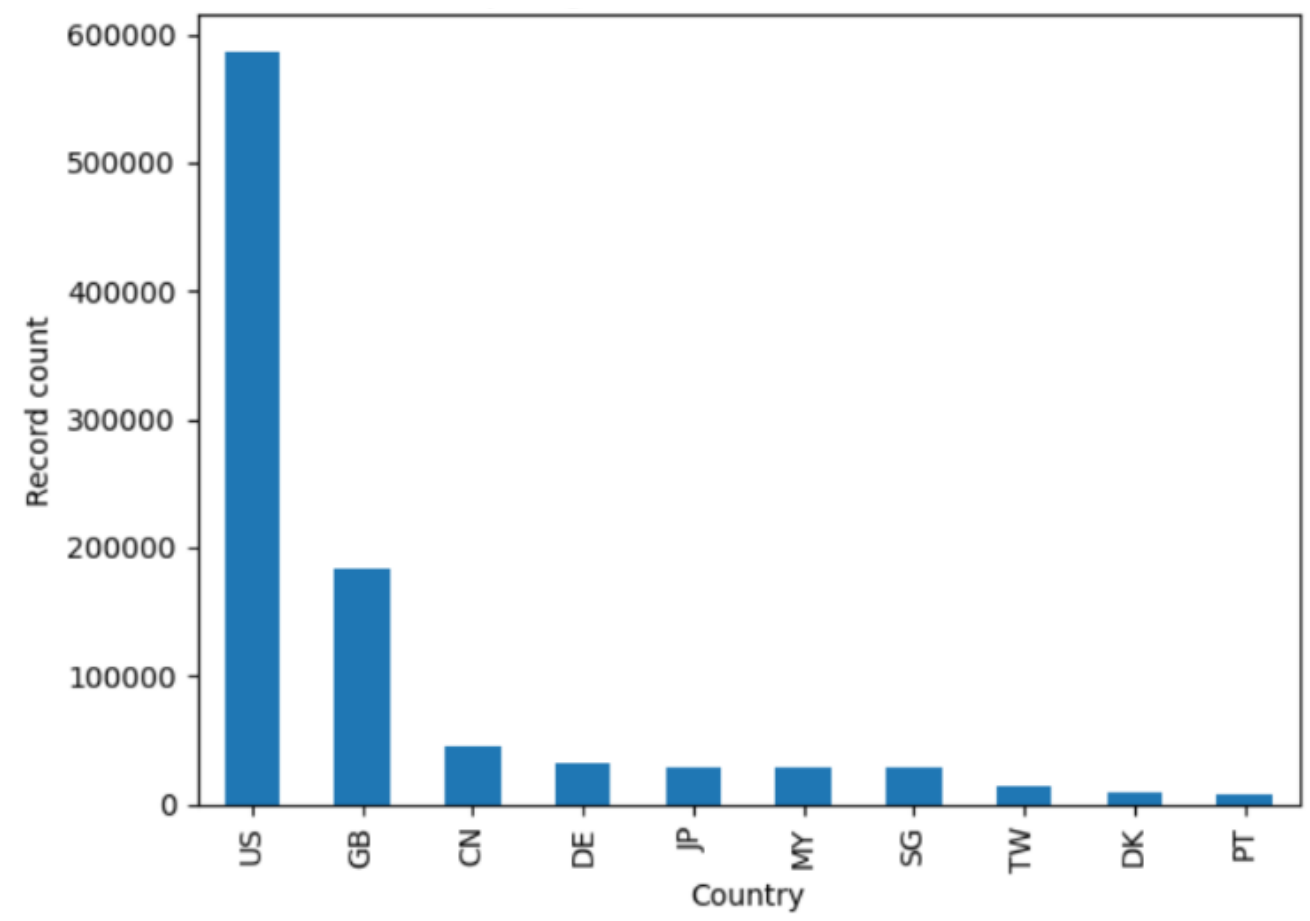

Fig. 3: Top 10 countries contributing unsolicited traffic.

### *C. ASN Volume and Burstiness Characteristics*

To further differentiate these source behaviors, Figure 5 classifies the top ten ASNs based on their total packet volume and temporal consistency. This multi-dimensional view confirms that the dataset is dominated by a heterogeneous ecosystem: **Persistent High-Volume Sources (Green):** AS214295 (Skynet) and AS201814 (Mevspace) are classified as persistent contributors. Despite fluctuations, their sustained presence across many active hours indicates a long-term reconnaissance mandate. **High-Volume Bursty Sources (Deep Red):** AS269538 (Conectec Net), AS396982 (Google), and AS8075 (Microsoft) exhibit extreme total volumes (over $10^6$ packets) but low temporal consistency. These represent the primary engines behind sudden traffic surges observed in the telescope. **Episodic Minor Contributors (Orange/Red):** Lower-ranked ASNs such as AS398324 (Amazon), AS16509 (Amazon), and AS401120 (Cloudflare) show moderate volume with high burstiness. This suggests the opportunistic use of transient cloud infrastructure for reconnaissance.The synchronization of bursty ASNs during high-traffic events, as seen in the late-window surge, underscores the transition from individual "background noise" to organized, distributed probing activities.

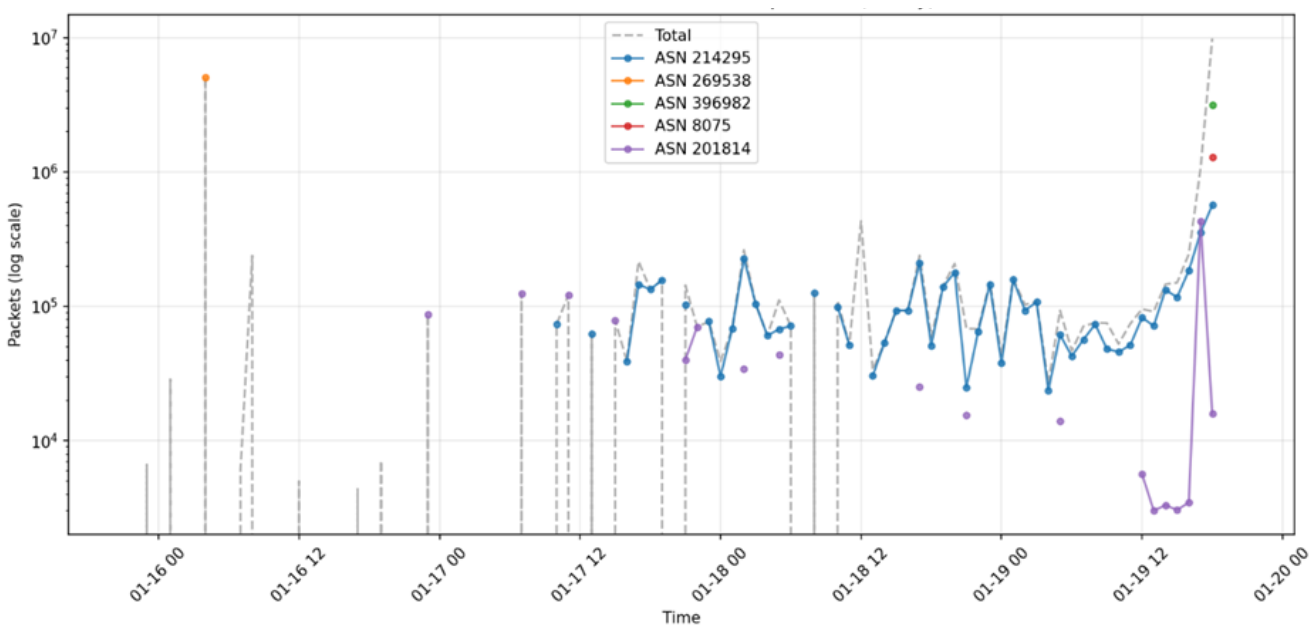

Fig. 4: Hourly traffic volume from the top five ASNs.

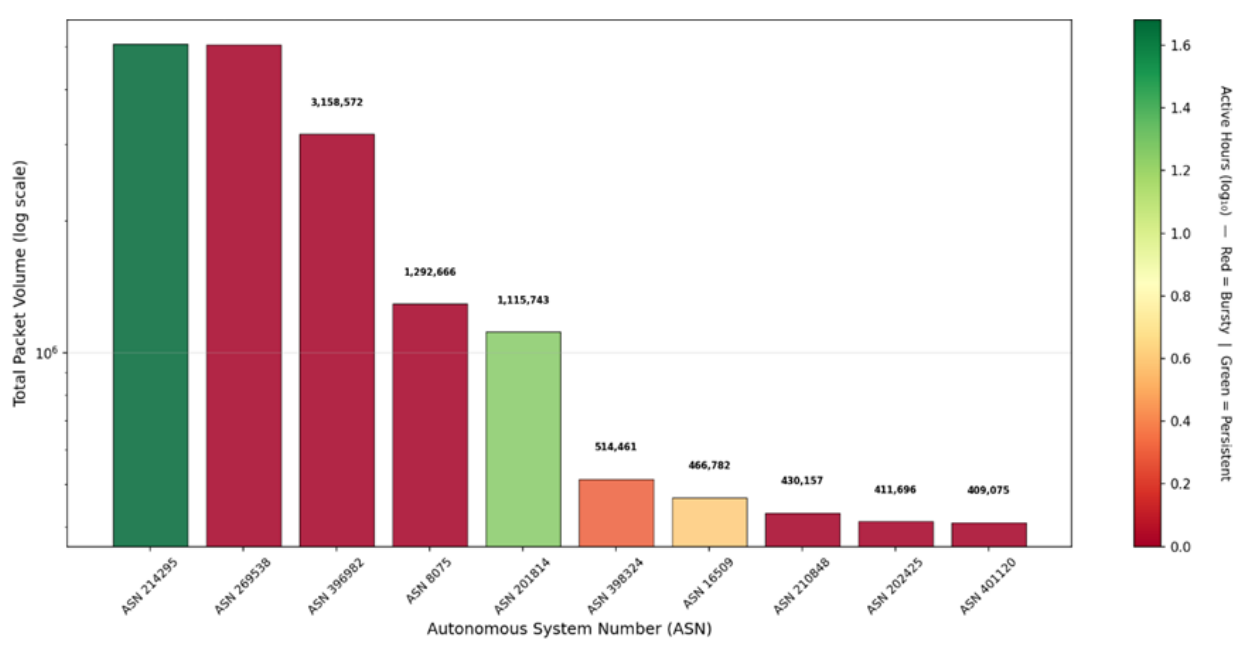

Fig. 5: Volume versus burstiness classification for the top ten ASNs. Inclusion of additional contributors reveals intermediate scanning patterns.

### D. High-Risk vs. Low-Risk Port Activity

Figure 6 compares traffic targeting high-risk services against ports classified in the lowest-risk quartile. The logarithmic scale highlights a dramatic disparity in engagement, confirming that unsolicited activity is highly selective.

The top five high-risk ports: 8080 (HTTP Alternate), 23 (Telnet), 80 (HTTP), 445 (Microsoft-DS/SMB), and 2323 (Telnet Alternate), account for majority of observed record counts. This concentration is indicative of specific scanning objectives: **Legacy and Alternate Administration:** The heavy presence of ports 23 and 2323 suggest widespread attempts to exploit weak credentials in legacy IoT devices and telnet-enabled hardware. **Web and File Sharing:** High volumes on ports 8080, 80, and 443 (HTTPS) reflect automated reconnaissance for vulnerable web applications, while activity on 445 aligns with known worm-like propagation tactics targeting the Server Message Block (SMB) protocol. **Specialized Services:** Notable traffic is also observed on ports 8443 (HTTPS Alt), 53 (DNS), 81 (Hosts2-ns), and 56766, suggesting a secondary focus on discovery and specialized service exploitation. Low-risk ports (e.g., 10146, 28174, and 62464) receive negligible traffic, often in the single-digit record counts. This extreme skew strongly indicates intentional reconnaissance rather than random port exploration.

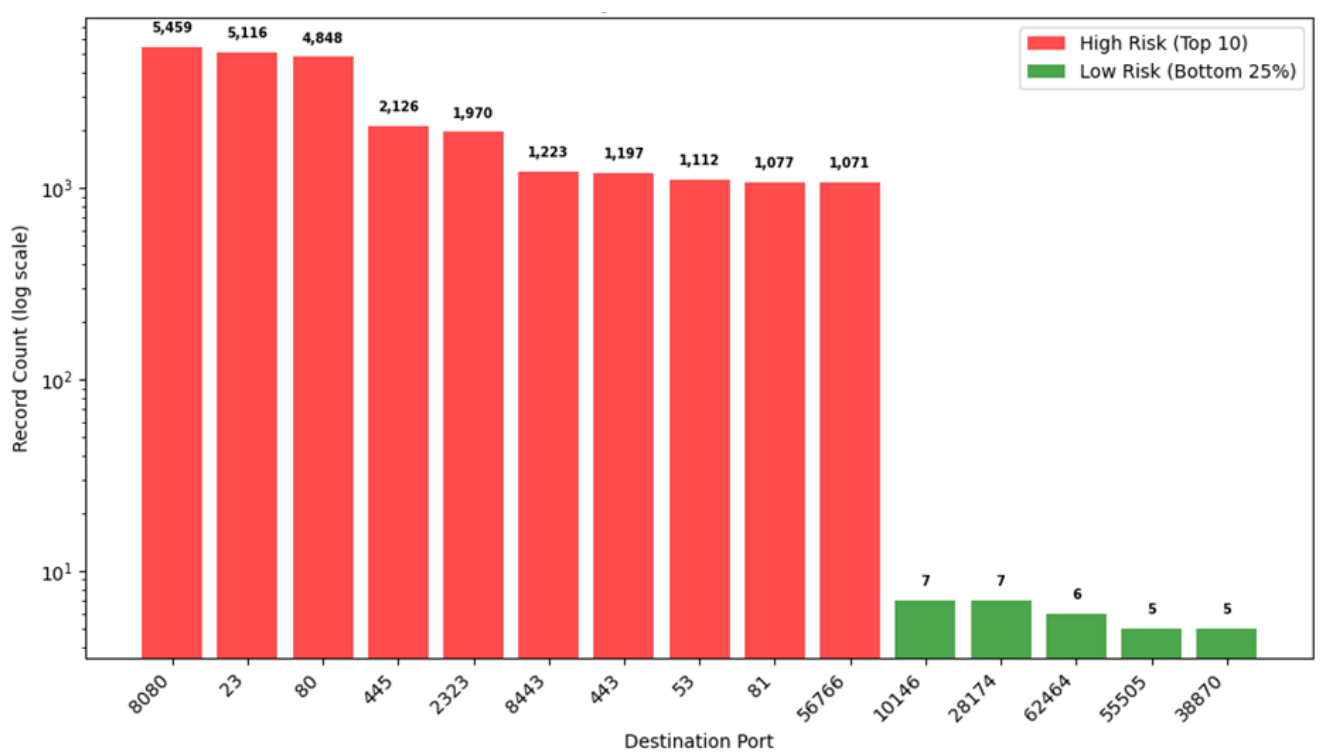

Fig. 6: Comparison of packet volume across high-risk and low-risk ports (log scale). High-risk services receive orders-of-magnitude more traffic than low-risk services.

### E. Traffic Concentration Analysis

To evaluate the distribution of unsolicited activity among the observed source population, we utilize a Lorenz curve as shown in Figure 7. The curve is sharply bowed toward the lower-right quadrant, deviating significantly from the line of perfect equality. This distribution results in a Gini coefficient of 0.981, which indicates a high degree of statistical inequality in traffic volume.

The data shows a highly concentrated distribution where a small minority of active sources accounts for the majority of the telescope's traffic. As summarized in Table III, the concentration of traffic is most dense within the top percentiles of source IP addresses.

TABLE III: Traffic concentration among top source IPs.

| Source Percentile | Unique IPs | Packet Volume | Cumulative |
|---|---|---|---|
| Top 1% | 392 | 17,882,842 | 81.19% |
| Top 5% | 1,959 | 20,953,536 | 95.13% |
| Top 10% | 3,918 | 21,823,916 | 99.08% |

The quantitative breakdown confirms that the top 1% of source IPs generate over 81% of total traffic, while the top 10% account for more than 99% of the observed records. Beyond the 10% threshold, the remaining 90% of the source

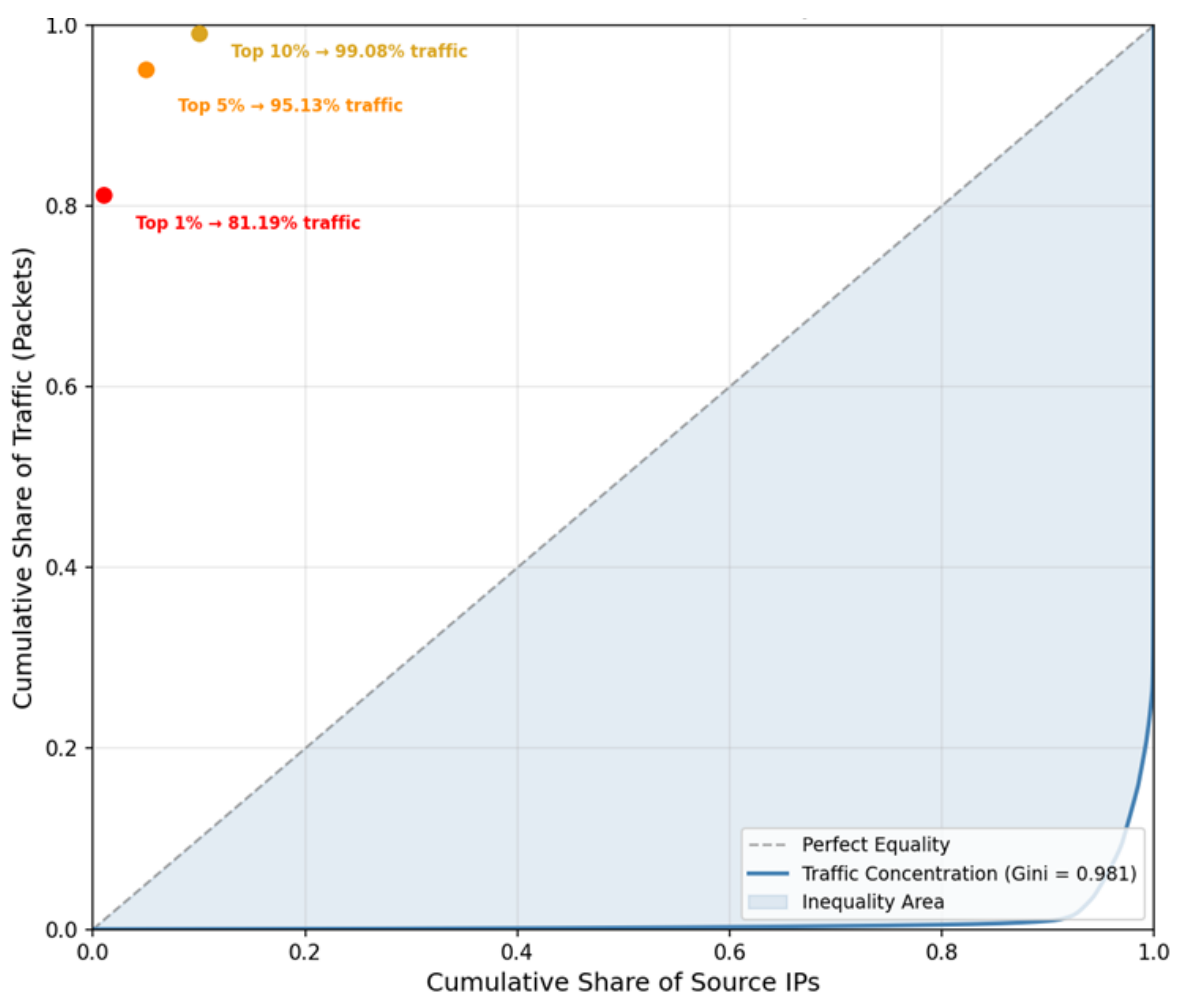

Fig. 7: Lorenz curve illustrating traffic concentration across source IP addresses. The Gini coefficient of 0.981 indicates extreme inequality in traffic distribution.

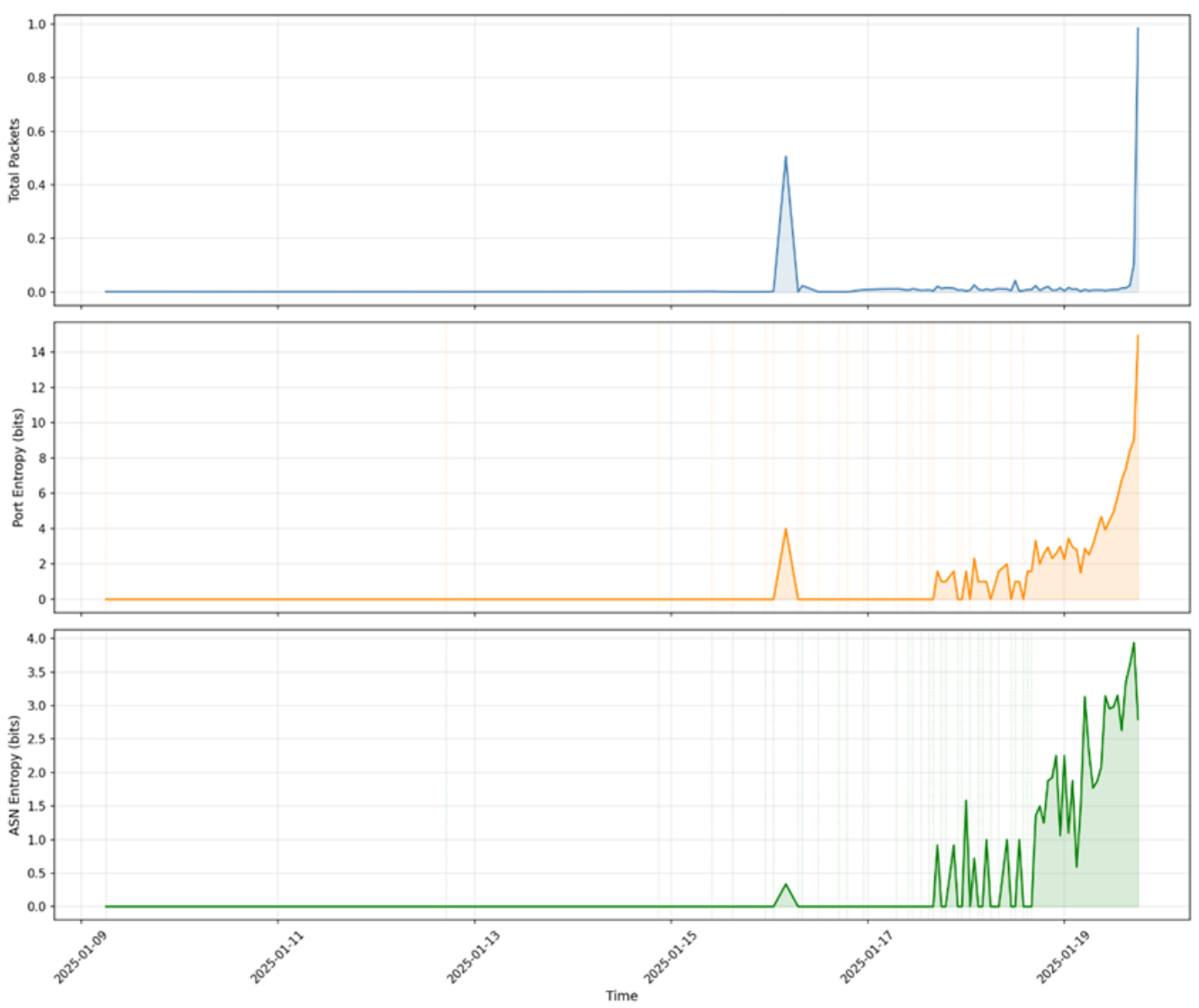

Fig. 8: Hourly traffic volume (top), raw port entropy (middle), and raw ASN entropy (bottom). Increases in entropy coincide with traffic surges.

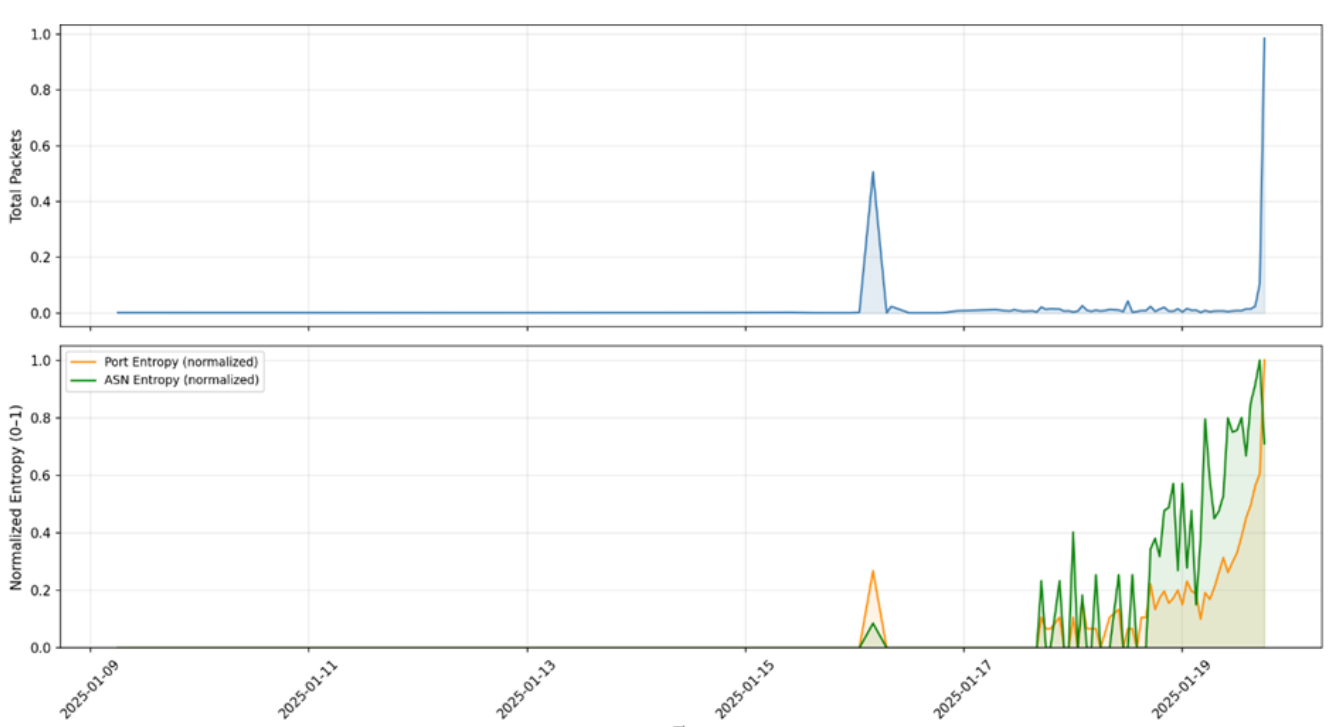

Fig. 9: Normalized entropy measures (0–1 scale) aligned with traffic volume. Normalization highlights synchronized increases in traffic and diversity.

population contributes less than 1% of the total packet volume.

This high degree of concentration suggests that the unsolicited traffic observed by the ORION telescope is not composed of random, decentralized background noise. Instead, it is characteristic of centralized or coordinated infrastructure, such as large-scale Internet measurement platforms, centralized botnet command-and-control nodes, or automated reconnaissance frameworks. These systems typically use a limited number of high-volume vantage points to achieve global coverage, leading to the extreme distribution patterns observed in this dataset.

### *F. Entropy-Based Diversity Analysis*

To characterize the structural complexity of the observed traffic, we analyze temporal diversity using Shannon entropy. Figure 8 illustrates the relationship between aggregate traffic volume and raw entropy values for both destination ports and source ASNs. Throughout the majority of the observation window, traffic remains negligible with near-zero entropy. However, a significant phase shift occurs between January 18 and January 20.

During this period, raw port entropy escalates to over 14 bits, while ASN entropy reaches approximately 4 bits. This simultaneous increase indicates that the observed traffic spikes are not the result of a single botnet targeting a lone service, but rather a diversified expansion in both the origins and targets of the unsolicited probes.

To confirm the correlation between traffic intensity and structural diversity, Figure 9 presents normalized entropy values. Scaling the metrics to a 0–1 range facilitates a direct comparison of their temporal signatures. The visualization reveals a highly synchronized rise: as packet volume reaches its peak on January 19, both ASN and port entropy achieve their maximum values.This synchronization suggests that large-scale traffic events in the ORION telescope are driven by distributed reconnaissance campaigns. Unlike localized probing, which would typically see volume increase while entropy stays low, these results highlight a coordinated expansion. The high ASN entropy suggests a broad, distributed source base, while the high port entropy reflects a multi-vector scanning strategy, targeting a wide array of services simultaneously. This pattern is consistent with modern, industrial-scale scanning ecosystems that optimize for breadth and speed across the IPv4 address space.

### *G. Discussion, Recommendations, and Future Work*

The observed patterns indicate that unsolicited Internet activity is far from random background noise: the strong concentration of traffic among a small subset of source networks implies large-scale probing is largely driven by centralized or coordinated infrastructure, and the contin-ued prominence of legacy IoT ports reflects long-standing weaknesses in authentication and device configuration.

Our behavioral heuristics, while lightweight and privacy-preserving, are not validated against ground-truth labeled data. Payload-equipped telescopes afford stronger validation: García-Peñas *et al.* [7] attributed ~89% of TCP traffic to ZMap, Masscan, or Mirai via packet-level signatures, and Cabana *et al.* [15] identified 13 distinct ICS payload templates via custom DPI on 12.85 TB of darknet traffic—forms of intent-level ground truth unavailable to metadata-only analyses. Irwin [16] similarly noted that IDS-based analysis using Snort/Bro was hampered by the absence of TCP payloads, a constraint identical to ours. Within the metadata-centric paradigm, Ismail *et al.* [1] provides the closest comparator, applying supervised ML to labeled ORION subsets; the precision and recall of our five-port scanning threshold relative to that benchmark remains to be formally established. Nevertheless, our structural findings—extreme concentration (Gini = 0.981) and Telnet/HTTP dominance—align closely with results from independent telescopes, providing indirect corroboration.

These findings translate into concrete recommendations. For **IoT manufacturers**, the persistent dominance of Telnet (ports 23 and 2323) calls for disabling Telnet by default in favor of SSH, enforcing per-device unique credentials to prevent Mirai-class credential-stuffing, and shipping automatic firmware updates given attackers' ability to scan the full IPv4 space within 24–48 hours of a CVE disclosure [5]. For **network operators**, the observation that 392 source IPs generate over 81% of telescope traffic suggests ASN-level block lists can meaningfully reduce exposure, while hourly Shannon entropy—which synchronized with the January 19 surge in our dataset—offers a payload-free early-warning signal feasible even for resource-constrained SOCs. Threat feeds should be weighted by ASN behavior rather than geography alone, since source geography largely reflects network topology rather than attacker location.

Future work will incorporate threat intelligence and vulnerability databases to correlate scanning activity with specific CVEs and exploit campaigns, apply machine learning to darknet metadata for automated source classification, and expand the observation window across additional telescope vantage points.

## V. Conclusion

The characterization of the ORION network telescope dataset demonstrates that unsolicited Internet traffic is a highly structured phenomenon dominated by centralized reconnaissance infrastructures. Results reveal extreme traffic concentration, with the top 1% of source IP addresses—primarily originating from the US and major cloud providers—generating over 81% of total traffic. The observed synchronization between traffic surges and peak Shannon entropy indicates that modern background radiation is shaped by coordinated, multi-vector campaigns targeting high-risk services like Telnet and HTTP to identify vulnerable IoT devices.

## Acknowledgment

This research was partially funded by NSF under the CICI: TCR program, IRIS: Instrumentation for Research and Inter-institutional SOC, Award Number NSF 2319793.